\documentclass[
               ,final            
               ]{aipproc}

\layoutstyle{8x11double}

\begin{document}

\title{Searching For New Thermally Emitting\\
       Isolated Neutron Stars In The 2XMMp Catalogue\\
       Discovery of a Promising Candidate}


\author{Adriana M. Pires}{
        address={Observatoire Astronomique, UMR 7550 CNRS, 11 rue de l'Universit\'e, 67000 Strasbourg, France}
        ,altaddress={Instituto de Astronomia, Geof\'isica e Ci\^encias Atmosf\'ericas, Universidade de S\~ao Paulo, R. do Mat\~ao 1226, 05508-090 S\~ao Paulo, Brazil}
}

\author{Christian Motch}{
        address={Observatoire Astronomique, UMR 7550 CNRS, 11 rue de l'Universit\'e, 67000 Strasbourg, France}
}

\begin{abstract}
The group of 7 thermally emitting and radio-quiet isolated neutron stars (INSs) discovered by ROSAT constitutes a nearby population which locally appears to be as numerous as that of the classical radio pulsars. So far, attempts to enlarge this particular group of INSs finding more remote objects failed to confirm any candidate. We found in the 2XMMp catalogue a handful of sources with no catalogued counterparts and with X-ray spectra similar to those of the ROSAT discovered INSs, but seen at larger distances and thus undergoing higher interstellar absorptions. In order to rule out alternative identifications such as an AGN or a CV, we obtained deep ESO-VLT and SOAR optical imaging for the X-ray brightest candidates. We report here on the current status of our search and discuss the possible nature of our candidates. We focus particularly on the X-ray brightest source of our sample, 2XMM J104608.7-594306, observed serendipitously over more than four years by the XMM-Newton Observatory. A lower limit on the X-ray to optical flux ratio of $\sim$ 300 together with a stable flux and soft X-ray spectrum make it the most promising thermally emitting INS candidate. Beyond the finding of new members, our study aims at constraining the space density of this population at large distances and at determining whether their apparently high local density is an anomaly or not.
\end{abstract}

\maketitle

\section{Introduction}
One of the outstanding results of ROSAT is the discovery of seven X-ray bright isolated neutron stars (INSs). These slowly rotating and radio-quiet neutron stars display thermal emission with $kT \sim$ 40 -- 110 eV, undergo little interstellar absorptions and are not associated with any supernova remnant (see reviews in \cite{tre00} and \cite{hab07}). Several have identified faint optical counterparts with $B \sim$ 25.8 -- 28.6. Proper motion studies (see Motch et al., these proceedings, and references therein) have shown that they are most probably young cooling neutron stars, with ages of a few 10$^5$ years. Their proximity and the apparent absence of strong non-thermal activity turn them into unique laboratories for testing radiative properties of neutron star surfaces in extreme conditions of gravitational and magnetic fields. Moreover, the possibility of measuring their distances through parallaxes \cite{ker07} or from the distribution of absorption on the line of sight \cite{pos07} can eventually bring important constraints on the debated equation of state of matter in neutron star interiors.

In the solar neighborhood, ROSAT INSs are as numerous as young radio and $\gamma$-ray pulsars. It is not clear whether this group is homogeneous. In particular, the absence of radio emission can be either due to the presence of intense magnetic fields, indeed inferred from the measured spin down rates and cyclotron X-ray spectral features, or due to the fact that the radio pencil beam, which narrows at long spin periods, does not sweep over the earth. Considering that ROSAT had not enough sensitivity and spatial resolution to detect the thermal emission of distant sources, the population of cooling radio-quiet INSs could represent a considerable fraction of the total neutron star population of the Galaxy, undetectable in radio surveys \cite{mot07}. In any case, our knowledge of the overall population characteristics will remain highly unsatisfactory as long as only seven objects are known.

\begin{table}
\begin{tabular}{l c r r c r r} 
\hline
\tablehead{1}{l}{t}{Source}     & 
\tablehead{1}{c}{t}{Cand}       & 
\tablehead{1}{r}{t}{RA}         & 
\tablehead{1}{r}{t}{DEC}        & 
\tablehead{1}{c}{t}{$R_{90}$}   &
\tablehead{1}{r}{t}{Count Rate} &
\tablehead{1}{r}{t}{Mag}\\
 & & (J2000) & (J2000) & arcsec & s$^{-1}$ & $R$\\
\hline
2XMM J104608.7-594306 & 065 & 10 46 08.7 & -59 43 06.1 & 1.33 & 0.060(4)   & $>$ 25\\ 
2XMM J121017.0-464609 & 164 & 12 10 17.1 & -46 46 11.2 & 2.60 & 0.027(6)   & 20.3  \\ 
2XMM J010642.3+005032 & 318 & 01 06 42.4 & +00 50 31.3 & 3.80 & 0.020(5)   & 24.5  \\ 
2XMM J214026.1-233222 & 364 & 21 40 26.2 & -23 32 22.3 & 1.90 & 0.0181(20) & 23.8  \\ 
2XMM J125904.5-040503 & 604 & 12 59 04.6 & -04 05 02.3 & 1.90 & 0.0129(21) & 21.2  \\ 
2XMM J125045.7-233349 & 681 & 12 50 45.7 & -23 33 47.7 & 3.30 & 0.0122(21) & 22.1  \\ 
\hline
\end{tabular}
\caption{INS candidates selected for optical investigation which have already been observed. Count rates are for the EPIC pn camera in the full XMM-Newton energy band (0.2--12.0 keV). We list the $R$ magnitude of the brightest object present in the error circles of the X-ray sources.\label{tab_cand}}
\end{table}

\begin{figure}
\includegraphics[height=.28\textheight]{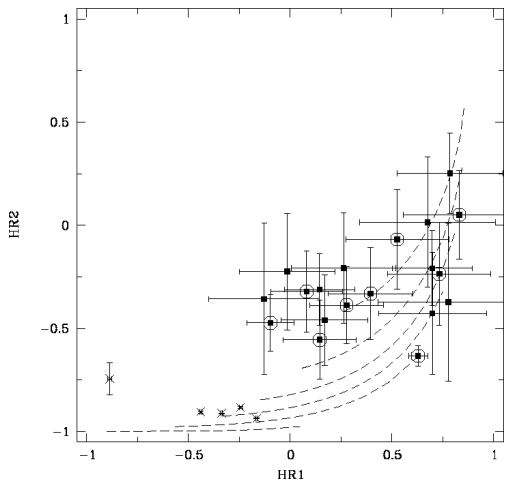}
\caption{The positions of our INS candidates, selected from $\sim$ 7.5$\cdot$10$^4$ XMM sources, are shown in the HR$_1 \times$ HR$_2$ diagram (squares). The known ROSAT INSs (crosses) occupy the lowest (less absorbed) part of the diagram. Dashed lines denote soft absorbed blackbodies of different temperatures (50-200 eV) and column absorptions. The nine INS candidates selected for optical follow-up during the present year are highlighted with circles. The X-ray brightest and most promising candidate (source 65) can be noticed by its somewhat smaller error bars. \label{fig_bestcand}}
\end{figure}

\section{Selection of the candidates}
In order to find more distant thermally emitting INSs, we searched the 2XMMp catalogue for new candidates using as criterion the absence of any catalogued optical object (USNO-A2, USNO-B1, SDSS) in the 3$\sigma$ error circles and hardness ratios all consistent with blackbodies with $kT \le$ 200 eV undergoing absorption columns in the range of 10$^{19}$\,--\,10$^{22}$\,cm$^{-2}$. We also checked that the X-ray absorption was less than or equal to the total galactic value in order to discard intrinsically absorbed sources. We only considered non-extended and well detected (maximum likelihood $\geq$ 8) sources with count rates brighter than $\sim$ 0.01 s$^{-1}$ in the EPIC pn camera.

For each selected source, we visually checked the X-ray and optical images and searched for possible identifications in over 170 astronomical catalogues. Many soft X-ray sources turned out to be false detections in extended diffuse emission or due to out of time events. Out of 7.5$\cdot$10$^4$ serendipitous sources above the mentioned flux, we found no more than $\sim$ 20 \emph{good} candidates -- i.e. intrinsically soft candidates not associated to any catalogued optical object. In Figure \ref{fig_bestcand}, are shown the positions of these sources in the hardness ratio diagram. The lowest left part of this diagram is occupied by the soft, low-absorbed ``Magnificent Seven'' sources while our candidates, undergoing higher photoelectric absorptions, move upwards in this diagram, along with the blackbody lines of hotter temperatures.

\section{Optical investigations}
We started an optical campaign aiming at the identification of the X-ray brightest candidates. The immediate objective is to discard a possible alternative identification, such as an extreme kind of CV or an AGN. Such objects can be easily identified from their colour indices and spectra. We have granted SOAR and ESO-VLT time to observe nine of our candidates.

\begin{table}
\begin{tabular}{l l c c c r r} 
\hline
\tablehead{1}{l}{t}{OBSID}               & 
\tablehead{1}{l}{t}{Detector}            & 
\tablehead{1}{c}{t}{$N_{\textrm{H}}$}    & 
\tablehead{1}{c}{t}{$kT$}                & 
\tablehead{1}{c}{t}{Flux}                &
\tablehead{1}{r}{t}{$\chi^2_{\nu}$}      &
\tablehead{1}{r}{t}{d.o.f.}\\
 & & 10$^{21}$ cm$^{-2}$ & eV & 10$^{-13}$ erg s$^{-1}$ cm$^{-2}$ & &\\
\hline
112560101 & pn       & 6.94$_{-0.20}^{+0.29}$ &  87$_{-18}^{+16}$ & 0.74$_{-0.06}^{+0.06}$ & 0.75 & 12\\
112580701 & pn M1 M2 & 1.52$_{-0.12}^{+0.26}$ & 169$_{-49}^{+45}$ & 0.92$_{-0.07}^{+0.07}$ & 0.73 & 25\\ 
145740201 & M1 M2    & 3.29$_{-0.17}^{+0.41}$ & 147$_{-51}^{+47}$ & 1.17$_{-0.12}^{+0.12}$ & 0.86 & 17\\
145740501 & M1 M2    & 6.64$_{-0.37}^{+0.24}$ &  99$_{-16}^{+43}$ & 0.93$_{-0.08}^{+0.12}$ & 0.66 & 17\\
145780101 & M1 M2    & 0.30$_{-0.03}^{+0.40}$ & 191$_{-86}^{+56}$ & 1.22$_{-0.14}^{+0.16}$ & 0.92 & 14\\
160160901 & M1 M2    & 2.87$_{-0.09}^{+0.15}$ & 132$_{-18}^{+23}$ & 0.96$_{-0.06}^{+0.06}$ & 0.81 & 18\\
\hline
\end{tabular}
\caption{Blackbody fit results for the best XMM-Newton observations of source 65. All errors are 1$\sigma$.\label{tab_X65}}
\end{table}

\begin{figure}
\includegraphics[height=.31\textheight]{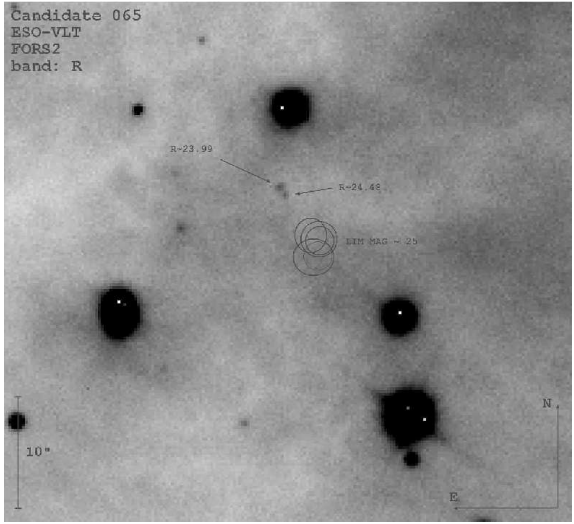}
\caption{Optical $R$ image of the field of candidate 65. The optical counterpart of the X-ray source is fainter than $R \sim$ 25. Circles are the 90\% confidence error on the position of the X-ray source as detected in several different XMM-Newton observations. The magnitudes of the two faint objects close to the X-ray source position are shown for comparison. Astrometry is based on the 2MASS catalogue.\label{fig_src65opt}}
\end{figure}

We list in Table \ref{tab_cand} the X-ray and optical information for each of the INS candidates which have already been observed. Preliminary analysis of our optical imaging data revealed the presence of faint ($R \sim$ 20.3 -- 24.5) optical objects in most X-ray error circles. We will use optical spectroscopy and colour indices to identify these objects and test their possible associations with the X-ray sources. Interestingly, our X-ray brightest candidate, source 2XMM J104608.7-594306 (or simply candidate 65), has no optical counterparts down to the limiting magnitude of our VLT image, $R \sim$ 25 (Figure \ref{fig_src65opt}).

\section{\centerline{X-ray properties of} \newline \centerline{2XMMU J104608.1-594306}}
We analysed the available X-ray archival data of source 65. Thanks to its proximity to the luminous Eta Carinae double star system, the source was serendipitously observed by the EPIC pn and MOS detectors on board XMM-Newton on several occasions, covering a total time span of more than 4 years. Unfortunately, in most of the observations, candidate 65 was located near the edge or in the CCD gaps and in many cases the exposure times were very short. We were however able to extract reasonably good spectra and carry out spectral analysis (see Table \ref{tab_X65}). The X-ray spectra are well fitted by a simple soft blackbody seen through a factor about 10 more interstellar absorption than for the ``Magnificent Seven''; no emission is detected above $\sim$ 1 keV and the flux does not seem to change very significantly over the 4 year time interval. The optical lower limit on the $R$ magnitude implies a $\log\big({f_{\textrm{x}}/f_{\textrm{opt}}}\big)$ ratio greater than 2.5 which definitely rules out a late type star identification and also probably excludes that the source is an extreme AM Her system. The high galactic extinction in that direction ($E_{(B-V)} \sim$ 12) rules out a background AGN. Preliminary timing analysis of the available data did not reveal the presence of coherent pulsations.

\section{Discussion}
Our search for new thermally-emitting INSs in the 2XMMp catalogue has revealed a handful of interesting and previously unknown soft X-ray sources among which source 65 is, by far, the most promising candidate. The analysis of its X-ray emission, although based on archival data obtained with non-optimal configurations, reveals an intrinsically soft energy distribution, apparently stable on long time scales. The derived $N_{\textrm{H}}$ is consistent with that observed towards Eta Carinae and its cluster ($N_{\textrm{H}} \sim$ 3$\cdot$10$^{21}$ cm$^{-2}$). Scaling from RX J0720.4$-$3125 yields a distance of $\sim$ 3.9 kpc probably compatible with the 2.5 kpc assumed for the Carina Nebula \cite{2007ApJ...656..462S}. Optical follow-up observations failed to reveal counterparts brighter than $R \sim$ 25, supporting the idea that source 65 is a new thermally emitting INS. 

We have already obtained optical data for some of the other candidates. Preliminary analysis of these data reveals the presence of faint optical candidates within the error circles in all cases. We plan to use optical colour indices and spectra to identify these sources, thus characterizing this sample of soft objects strictly selected from more than 120 thousand entries present in the 2XMMp catalogue.

\begin{theacknowledgments}
This work has been supported by Funda\c c\~ao de Amparo \`a Pesquisa do Estado de S\~ao Paulo (FAPESP), Coordena\c c\~ao de Aperfei\c coamento de Pessoal de N\'ivel Superior (CAPES), Brazil and by Universit\'e Louis Pasteur, Strasbourg, France.
\end{theacknowledgments}

\bibliographystyle{aipproc}

\bibliography{pires_adriana_1}

\end{document}